\documentclass[11pt,a4paper]{article} 
\usepackage[dvips]{graphicx}
\usepackage{epstopdf}
\usepackage{amsfonts}
\usepackage{amsthm}
\usepackage{amsmath}
\usepackage{epsfig}
\usepackage{t1enc}
\usepackage{amssymb}
\usepackage{latexsym}
\usepackage{bm}
\usepackage[hang]{caption}

\usepackage{ulem}

\usepackage{wrapfig}
 
\usepackage[extra]{tipa}

\def\tG{\mbox{\doubletilde{$G$}}} \def\tx{\mbox{\doubletilde{$\xi$}}}
\def\te{\mbox{\doubletilde{$\eta$}}} \def\tte{\mbox{\scriptsize\doubletilde{$\eta$}}}

\title{\bf Constraints on modified dispersion relations}

\author{Vladim\'ir Balek\footnote{e-mail
address: balek@fmph.uniba.sk}\ \ and Orchidea Maria Lecian
\footnote{e-mail address: omlecian@gmail.com}
\\
{\it Department of Theoretical Physics, Comenius University,
Bratislava, Slovakia}}

\begin{document}

\renewcommand{\figurename}{\small Fig.}

\maketitle \maketitle\abstract

{Deviation from standard dispersion relations for electrons and
photons in the form of an extra term proportional to an
arbitrarily high power of momentum is studied. It is shown that
observational constraints lead to a region in the parametric space
that is similar in shape to the region obtained earlier in a
theory in which the extra term was proportional to third power of
momentum.}

\section{Introduction}

Parallel to exploring possible paths to unification of quantum
mechanics with general relativity within a well-established
theoretical framework, like string theory or loop quantum gravity,
there exists an approach in which ``simple (in some cases even
simple-minded) non-classical pictures of spacetime are being
analyzed with strong emphasis on their observable predictions''
\cite{ac}. In this approach, known as {\it quantum--gravity
phenomenology}, one seeks effects observable with present--day
experimental devices in order to sort out existing concepts and
ideas in quantum--gravity and provide guidance for developing new
ones; for a review, see \cite{acrev}. An important ingredient of
quantum--gravity phenomenology is investigation of the
consequences which would appear if the particles acquired, due to
interaction with spacetime foam, a small additional term in
dispersion relation. The idea was proposed in \cite{{acell},{col}}
and was analyzed, for the extra term proportional to $n$-th power
of momentum, in detail in \cite{jac}. As for later developments,
loop theory implications are examined in \cite{{bojo},{gir}},
connection with generalized uncertainty principle is discussed in
\cite{spr} and new observational data are reviewed in
\cite{{lib},{mar},{anch}}. Furthermore, in
\cite{{pfs},{fpfs},{gro}} it is explored how modified dispersion
relations arise in electrodynamics in media, and in
\cite{{bbglph},{pfei}} it is examined how a possible modification
of dispersion relations would manifest itself in black hole
physics and cosmology.

\vskip 4mm In \cite{jac} the analysis was done for $n = 2, 3$ and
(partly) 4, here we extend it to arbitrarily large $n$. We
restrict ourselves to three processes which were shown in
\cite{jac} to be crucial for determining the allowed region in the
parametric space: vacuum \v Cerenkov radiation, photon decay and
collision of two photons with creation of an electron-positron
pair. In section 2 we analyze the first two processes, in section
3 we investigate the third process and in section 4 we discuss
results.

\section{One-particle processes}

Consider dispersion relations for photon (energy $\omega$,
momentum $\bf k$) and electron (energy $E$, momentum $\bf p$, mass
$m$) of the form
\begin{equation}
\omega^2 = k^2 + \xi k^n, \quad E^2 = m^2 + p^2 + \eta p^n,
\label{1}
\end{equation}
where $\xi$ and $\eta$ are parameters with physical dimension
mass$^{-(n - 2)}$. (We are using units in which $c = 1$.) As it
turns out, for $n = 2$ the results are qualitatively different
than for other values of $n$, and since this value was discussed
in detail in \cite{jac}, we skip it from the analysis and restrict
ourselves to $n \ge 3$.

\vskip 4mm With modified dispersion relations, we have to include
into the theory two processes with one particle in the initial
state, which are normally forbidden due to energy-momentum
conservation: {\it vacuum \v Cerenkov radiation} and {\it photon
decay}
\begin{figure}[ht]
\centerline{\includegraphics[height=3cm]{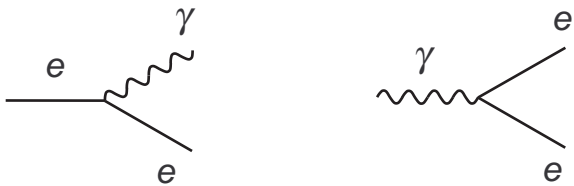}}
\caption{\small One-particle processes}
\end{figure}
(fig. 1). Denote the 4-momentum of the photon by \uuline{$k$} and
the 4-momenta of the remaining two particles (in- and outgoing
electron in the first process and electron and positron in the
second process) by \uuline{$p$} and \uuline{$q$}. We have
$\mbox{\uuline{$p$}} = \mbox{\uuline{$k$}} + \mbox{\uuline{$q$}}$
for the first process and $\mbox{\uuline{$k$}} =
\mbox{\uuline{$p$}} + \mbox{\uuline{$q$}}$ for the second process,
so that for both processes the 4-momenta satisfy
$(\mbox{\uuline{$k$}} - \mbox{\uuline{$p$}})^2 =
\mbox{\uuline{$q$}}^2$. Suppose the electrons are
ultrarelativistic, $p$ as well as $q \gg m$. At the threshold,
where the momenta $\bf k$, $\bf p$ and $\bf q$ are parallel to
each other, the equation reduces to
$$\xi k^n \Big(1 - \frac pk\Big) + \eta \Big[p^n \Big(1 - \frac
kp\Big) - q^n\Big] = m^2 \frac kp,$$ where $q = p - k$ and $k - p$
for the first and second process respectively. Introduce
dimensionless variables $x = k/p$, $0 < x < 1$, for the first
process, and $y = p/k$, $0 < y < 1$, for the second process.
Rewritten in terms of $x$ and $y$, the equation for the first
process reads
\begin{equation*}
F \equiv \bar x [-\xi x^{n - 2} + \eta (1 + \bar x + \ldots + \bar
x^{n - 2})] = a, \tag{2a}
\end{equation*}
where $\bar x = 1 - x$ and $a = m^2/p^n$, and the equation for the
second process reads
\begin{equation*}
G \equiv  y\bar y [\xi - \eta (y^{n - 1} + \bar y^{n - 1})] = b,
\tag{2b}
\end{equation*}
where $\bar y = 1 - y$ and $b = m^2/k^n$.

\vskip 4mm Obviously, the first process can take place only if $F
> 0$ and the second process can take place only if $G > 0$. This
suggests that there are two regions in the $(\eta, \xi)$ plane,
one for each process, which are safe in the sense that the
processes cannot take place in them ($F \le 0$ for the first
process and $G \le 0$ for the second process), no matter what the
momentum of the incoming particle. The functions $F$ and $G$ can
be written as
$$F \propto - \xi f + \eta,\ f = x^{n -
2}/(1 + \bar x + \ldots + \bar x^{n - 2}), \mbox{ and } G \propto
\xi - \eta g,\ g = (y^{n - 1} + \bar y^{n - 1}),$$ and a simple
analysis shows that $f$ ranges from 0 to 1 (it rises monotonically
from 0 at $x = 0$ to 1 at $x = 1$), while $g$ ranges from $2^{-(n
- 2)}$ to 1 (it falls down from 1 at $y = 0$ to $2^{-(n - 2)}$ at
$y = 1/2$ and rises back to 1 at $y = 1$). As a result, the safe
region for the first process is
\begin{equation*}
\xi \ge 0: \eta \le 0 \quad \bigcup \quad \xi < 0: \eta \le \xi,
\tag{3a}
\end{equation*}
and the safe region for the second process is
\begin{equation}
\xi \ge 0: \eta \ge 2^{n - 2} \xi \quad \bigcup \quad \xi < 0:
\eta \ge \xi, \tag{3b}
\end{equation}
see fig. 2. Note that when both processes are taken into account,
one is left with the safe ray on the lower diagonal ($\eta = \xi
\le 0$) only.
\begin{figure}[ht]
\centerline{\includegraphics[height=3cm]{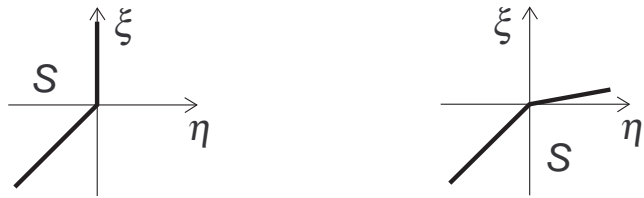}}
\caption{\small Safe regions}
\end{figure}

\setcounter{equation}{3}

\vskip 2mm The region in the $(\eta, \xi)$ plane which is allowed
by observations is given by the inequalities $F_{max} < A$ and
$G_{max} < B$, where $A$ and $B$ are observational upper bounds on
$a$ and $b$, defined in terms of maximum observed momenta of
electrons and photons coming from extragalactic sources
$p_{max}^{obs}$ and $k_{max}^{obs}$ as $A = m^2/(p_{max}^{obs})^n$
and $B = m^2/(k_{max}^{obs})^n$. The inequalities define a region
in the $(\eta, \xi)$ plane between the lines
\begin{equation}
F_{max} (x; \xi, \eta) = A, \quad G_{max} (y; \xi, \eta) = B.
\end{equation}
Our goal is to find how this region looks like for arbitrary $n$.

\vskip 4mm Write the functions $F$ and $G$ as
$$F = \bar f ( - \xi f + \eta),\ \bar f = \bar x (1 + \bar x + \ldots
+ \bar x^{n - 2}), \mbox{ and } G = \bar g(\xi - \eta g),\ \bar g
= y\bar y.$$ We can easily see that the functions $\bar f$ and
$\bar g$ behave complementary to the functions $f$ and $g$: when
the latter functions rise, the former functions fall, and {\it
vice versa}. Specifically, $\bar f$ falls monotonically (it starts
from $n - 1$ at $x = 0$ and falls to 0 at $x = 1$) and $\bar g$
first rises and then falls (it starts from 0 at $y = 0$, rises to
$1/4$ at $y = 1/2$ and falls back to 0 at $y = 1$). Thus, if $F$
and $G$ are positive, which is the case we are interested in, the
function $F$ falls monotonically if $\xi > 0$ and the function $G$
first rises and then falls if $\eta > 0$. This suggests that the
maximum values of $F$ and $G$ are
$$\mbox{$\xi > 0$:}\ F_{max} = F(0) = (n - 1)\eta, \quad
\mbox{$\eta > 0$:}\ G_{max} = G(1/2) = (1/4)(\xi - 2^{-(n -
2)}\eta),$$ and the allowed region in the first quadrant of
$(\eta, \xi)$ plane is a strip adjacent to the axes, delimited
from the right and from above by the straight lines
\begin{equation}
\mbox{$\xi > 0$:}\ \eta = A/(n - 1) \equiv \eta_0, \quad
\mbox{$\eta > 0$:}\ \xi = 4B + 2^{-(n - 2)}\eta \equiv \xi_0
(\eta).
\end{equation}

\vskip 2mm Consider now the function $F$ for $\xi < 0$ and the
function $G$ for $\eta < 0$. For such $\xi$ and $\eta$, the
functions ${\cal F} = - \xi f + \eta$ and ${\cal G} = \xi - \eta
g$ behave in the same way as the functions $f$ and $g$: the
function $\cal F$ rises monotonically and the function $\cal G$
first falls and then rises. Consequently, the behavior of the
functions $F$ and $G$ changes. Consider first the function $F$
with the parameter $\eta$ equal to $\eta_0$. If $\xi$ becomes
negative, the function acquires a bump whose height rises as $\xi$
decreases, and eventually, for $\xi$ equal to some critical value
$\xi_c$, it reaches the value $A$. If we continue to decrease
$\xi$, the height of the bump continues to increase, so that in
order to keep the maximum of $F$ equal to $A$, $\eta$ must start
to decrease. As a result, the line $\alpha$ delimiting the allowed
region from the right is vertical at $\xi > \xi_c$ (the
$\alpha^{(+)}$ part) and bent to the left at $\xi < \xi_c$ (the
$\alpha^{(-)}$ part). The behavior of the function $G$ in the
interval $1/2 \le y \le 1$ is similar, we just have to interchange
the variables $\xi$ and $\eta$. Thus, for $\xi = \xi_0$ the
function acquires a bump with increasing height, which for $\eta$
equal to some critical value $\eta_c$ crosses the value $B$, and
from that moment on $\xi$ must fall at a higher rate than $\xi_0$.
As a result, the line $\beta$ delimiting the allowed region from
above is straight, tilted downwards at $\eta > \eta_c$ (the
$\beta^{(+)}$ part) and bent downwards at $\eta < \eta_c$ (the
$\beta^{(-)}$ part). The behavior of the functions $F$ and $G$, as
well as the form of the allowed region, is depicted in fig.
\begin{figure}[ht]
\centerline{\includegraphics[height=5.2cm]{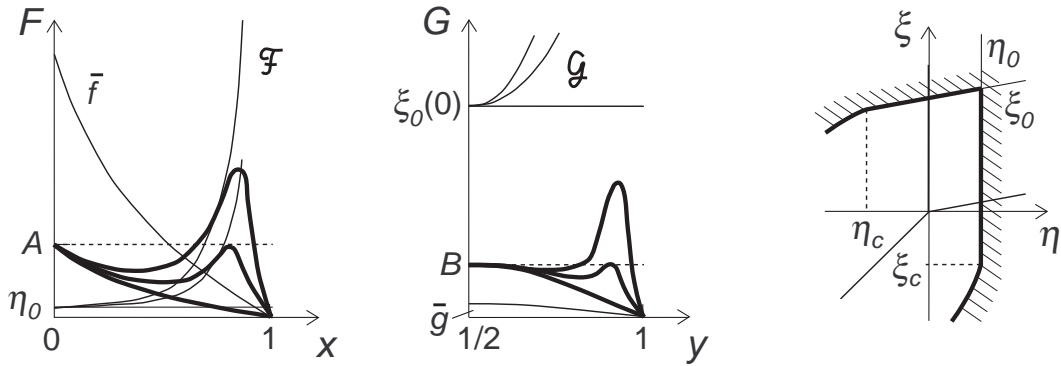}}
 \caption{\small Construction of the allowed region}
\end{figure}
3. The three heavy lines in the left panel are the graphs of the
function $F$ for $\eta = \eta_0$ and $\xi = 0$, $\xi = \xi_c$,
$\xi > \xi_c$, and the three heavy lines in the central panel are
the graphs of the function $G$ for $\xi = \xi_0$ and $\eta = 0$,
$\eta = \eta_c$, $\eta > \eta_c$. Light lines depict, as indicated
in the figure, functions $\bar f$, $\cal F$ and $\bar g$, $\cal G$
appearing in the expressions for $F$ and $G$. The resulting
allowed region in the $(\eta, \xi)$-diagram is the unshaded region
between the heavy lines in the right panel. Note that for $n = 3$
the bent parts of the boundary of the allowed region (the lines
$\alpha^{(-)}$ and $\beta^{(-)}$) match the adjacent straights
parts (the lines $\alpha^{(+)}$ and $\beta^{(+)}$) without
changing direction.

\vskip 4mm By analyzing the behavior of $F'$ one finds that for
$\xi < 0$ the function $F$ has just one bump which becomes local
maximum as $\xi$ crosses some value $\xi_{c1}> \xi_c$. If the
value of $\xi$ further decreases, first the local maximum of $F$
increases monotonically and then, after $\xi$ crosses the value
$\xi_c$, the maximum becomes global and stays constant provided
the value of $\eta$ decreases monotonically. The function $G$ for
$\eta < 0$ behaves analogically, if we restrict ourselves to $y$
running from 1/2 to 1. Let us verify the monotonic decrease of
$\eta$ with decreasing $\xi$ beyond the critical point and prove
in such a way that the maximum of $F$ is constant there. The
function $\eta (\xi)$ is given by the equations $F(x; \eta, \xi) =
A$ and $\partial_x F(x; \eta, \xi) = 0$. If we insert $F = \bar f
(- \xi f + \eta)$ into the first equation and differentiate it
with respect to $\xi$, we obtain
$$\bar f \Big(- f + \frac {d\eta}{d\xi}\Big) + \frac
{\partial [\bar f (- \xi f + \eta)]}{\partial x} \frac {dx}{d\xi}
= 0,$$ and if we use the second equation, we find that $d\eta/d\xi
= f > 0$. Analogically we can prove for the function $\xi (\eta)$
beyond the critical point that $d\xi/d\eta = g > 2^{-(n - 2)}$.
Thus, the lines $\alpha$ and $\beta$ are both bent towards the
lower diagonal beyond the critical point.

\vskip 4mm Let us determine $\xi_c$ and $\eta_c$ (longitudinal
shifts of the critical points with respect to the origin) for $n
\gg 1$. Denote the quantities rescaled by $A$ by a hat and the
quantities rescaled by $B$ by a tilde. The functions appearing in
the expression for $F$ are $\bar f f = \bar x x^{n - 2}$ and $\bar
f = \bar x (1 + \bar x + \ldots \bar x^{n - 2})$, and since $x$
turns out to be close to 1, for the latter function we have $\bar
f \doteq \bar x$. To find the quantity $\hat \xi_c$ we must solve
equations $F(x; \hat \eta_0, \hat \xi_c) = 1$ and $\partial_x F(x;
\hat \eta_0, \hat \xi_c) = 0$, where $\hat \eta_0 = 1/(n - 1)$.
The first equation yields $- \hat \xi_c \bar x x^{n - 2} \doteq 1
- \hat \eta_0 \bar x \doteq 1$, hence $\hat \xi_c \doteq - 1/(\bar
x x^{n - 2})$, and the second equation yields $- \hat \xi_c (\bar
x x^{n - 2})' = \hat \xi_c x^{n - 3} [1 - (n - 1)\bar x] \doteq
\hat \eta_0 \doteq 0$, hence $\bar x \doteq 1/(n - 1)$ and $x^{n -
2} \doteq [1 - 1/(n - 1)]^{n - 2} \doteq e^{-1}$. The resulting
expression for the quantity $\hat \xi_c$ is $\hat \xi_c \doteq -
e(n - 1)$, and if we perform an analogical calculation with the
function $G$, we obtain $\tilde \eta_c \doteq - e(n + 1)$. Of
course, the results are valid only in the leading order in
$n^{-1}$, therefore we can neglect $\pm 1$ in the brackets and
write
\begin{equation}
\hat \xi_c \mbox{ as well as } \tilde \eta_c \doteq -en
\end{equation}
We can also see that the critical points are much further from the
origin in the longitudinal direction than in the transversal
direction, $|\hat \xi_c| \gg \hat \eta_0 \doteq n^{-1}$ and
$|\tilde \eta_c| \gg \tilde \xi_0 (\eta_c) \doteq 4$.

\vskip 4mm Finally, let us determine the asymptotic form of the
lines $\alpha^{(-)}$ and $\beta^{(-)}$ far from the origin. We are
interested in the function $\eta (\xi)$ defined by the condition
$F_{max} = A$ and the function $\xi (\eta)$ defined by the
condition $G_{max} = B$ for $|\xi| \gg |\xi_c|$ and $|\eta| \gg
|\eta_c|$ respectively. Consider the former function. The value of
$x$ for which $F_{max} = A$ is now close to 1 for any $n$, and is
much closer to 1 than in the case $\xi = \xi_c$ for $n \gg 1$,
therefore we can write $\bar f f \doteq \bar x [1 - (n - 2) \bar
x]$. To the same order of magnitude, $\bar f \doteq \bar x (1 +
\bar x)$. (For $n = 3$, this is exact.) Consequently, for the
function $\hat F = F(x; \hat \eta, \hat \xi)$ we have $\hat F
\doteq (\hat \eta - \hat \xi) \bar x + [\hat \eta + (n - 2)\hat
\xi] \bar x^2$, and if we denote $\hat \eta_\pm = \hat \eta \pm
\hat \xi$ and use that, as seen from the final formula, $\hat
\eta_- \ll |\hat \eta_+|$, we can write
$$\hat F \doteq \hat \eta_- \bar x + \frac 12 (n - 1)\hat
\eta_+ \bar x^2.$$ Equation $\hat F' = 0$ yields $\bar x \doteq -
1/(n - 1) (\hat \eta_-/\hat \eta_+)$, and if we insert this into
the equation $\hat F = 1$, we find that the line $\alpha_{lim}$
approached by $\alpha^{(-)}$ far from the origin is given by
\begin{equation*}
\hat \eta_+ = - \frac {\hat \eta_-^2}{2(n - 1)}, \quad \hat \eta_-
> 0. \tag{7a}
\end{equation*}
In a similar manner we obtain the line $\beta_{lim}$. The formula
for it turns out to be the same as for the line $\alpha_{lim}$, we
just have to replace the quantities with a hat by the quantities
with a tilde and consider complementary definition region. Thus,
$\beta_{lim}$ is given by
\begin{equation*}
\tilde \eta_+ = - \frac {\tilde \eta_-^2}{2(n - 1)}, \quad \tilde
\eta_- < 0. \tag{7b} \label{adec}
\end{equation*}
We can see that the limit lines are halves of two parabolas with
the axis on the lower diagonal, whose widths are in general
different, but become identical for $A = B$
\begin{figure}[ht]
\centerline{\includegraphics[height=4.8cm]{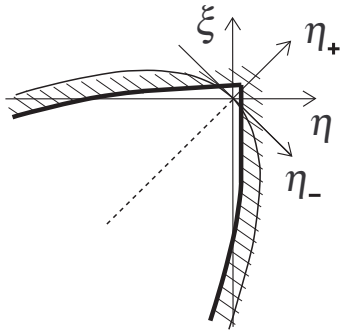}}
 \caption{\small Allowed region far from the origin}
\end{figure}
(fig. 4). The line $\alpha_{lim}$ is the lower half and the line
$\beta_{lim}$ is the upper half of the respective parabola.

\vskip 4mm The boundaries of the allowed region converge to the
limit lines in general only in a weak sense: they copy their
shape, but keep finite distance from them. For $n > 3$, the shifts
of the true limit lines along the axes $\eta_-$ and $\eta_+$ are
given by the expansion of the functions $\hat F$ and $\tilde G$ up
to third and fourth order respectively. For $n = 3$, the
boundaries coincide with the limit lines and are of the form $\hat
\eta_+ = -(1/4)\hat \eta_-^2$, $\hat \eta_- > 2$, and $\tilde \eta
= -(1/8) \tilde \eta_-^2$, $\tilde \eta_- < -8$; thus, the shifts
of the limit lines along the axes $\eta_-$ and $\eta_+$ are
$\Delta \hat \eta_\pm = 0$ and $\Delta \tilde \eta_- = -2$,
$\Delta \tilde \eta_+ = 1$.

\section{Two-particle process}

The two processes considered so far have left us with an allowed
region in the form of an infinite wedge around the lower diagonal
in the $(\eta, \xi)$ plane. To cut the region from below, let us
consider {\it collision of two photons},
 \begin{figure}[ht]
\centerline{\includegraphics[height=3.8cm]{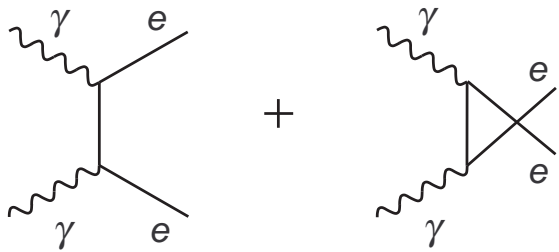}}
 \caption{\small Two-particle process}
\end{figure}
hard and soft, {\it with creation of an electron-positron pair}
(fig. 5). The process can take place, unlike the previous two,
also in a Lorentz invariant theory. However, after passing to a
theory with modified dispersion relations we find that the lower
threshold shifts in one direction or another, and there possibly
appears an upper threshold as well.

\vskip 4mm The constraint on particle momenta in the photon
collision is most easily obtained if we use the previous analysis
for photon decay with the replacements $\omega \to \omega_1 =
\omega + \omega_0$, $k \to k_1 = k - \omega_0$, where $\omega_0$
is the frequency of the soft photon. From the expression of
$\omega_1$ in terms of $k_1$ we recover the previous theory with
the replacement $\xi \to \xi_1 = \xi + 4\omega_0/k^{n_-}$, where
$n_- = n - 1$. Thus, the constraint we are looking for reduces to
the constraint for photon decay with an extra term on the left
hand side, \setcounter{equation}{7}
\begin{equation}
G_{col} \equiv  y\bar y [\xi - \eta (y^{n_-} + \bar y^{n_-}) +
4\omega_0/k^{n_-}] = b.
\end{equation}
The lower threshold for pair creation is defined as the minimum of
the variable $k$ given by this equation (which contains $k$ also
on the right hand side, since $b \propto k^{-n}$), provided $\xi$
and $\eta$ are fixed and $y$ is running from 0 to 1. Instead of
$k$ it is convenient to work with the dimensionless parameter
$\beta = k/k_{LI}$, where $k_{LI}$ is the threshold for pair
creation in a Lorentz invariant theory, $k_{LI} = m^2/\omega_0$.
The constraint on momenta expressed in terms of $\beta$ reads
\begin{equation}
\tG_{col} \equiv  y\bar y [\beta^n \tx - \beta^n \te (y^{n_-} +
\bar y^{n_-}) + 4\beta] = 1,
\end{equation}
where the double tilde denotes rescaling by the dimensional
constant $\omega_0^n/m^{2n_-}$. The function $\tG_{col} - 1$ is a
polynomial of $n$th order in $\beta$, therefore equation
$\tG_{col} = 1$ defines a function $\beta(y; \tx, \te)$ which may
have as many as $n$ real values for given $y$. The lower threshold
of $e^- e^+$ creation in the units $k_{LI}$ is the minimum of this
function when restricted to positive values.

\vskip 4mm Rewrite equation $\tG_{col} = 1$ as a definition of the
function $\tx (y; \beta, \te)$,
\begin{equation}
\tx = \te (y^{n_-} + \bar y^{n_-}) + \beta^{-n} [(y\bar y)^{-1} -
4\beta]. \label{tx}
\end{equation}
The parameter $\beta$ has an extremum as a function of $y$ if
$\beta' = -\partial_y \tx/\partial_\beta \tx = 0$. It holds
\begin{equation*}
\partial_y \tx = [n_- \te\phi + \beta^{-n} (y\bar
y)^{-2}] (y - \bar y), \quad \phi = y^{n - 3} +  y^{n - 4} \bar y
+ \ldots + \bar y^{n - 3},
\end{equation*}
therefore there exists always an extremum at $y = \bar y = 1/2$,
and for $\te < 0$ there may exist also pairs of extrema at $y <
1/2$ and $y > 1/2$, located symmetrically with respect to the
point $y = 1/2$. Thus, unlike in Lorentz invariant theory where
the threshold configuration is necessarily symmetric (has $y =
1/2$), in a theory with modified dispersion relations there may
exist also threshold configurations that are asymmetric. For
definiteness, we will suppose that these configurations have $y <
1/2$.

\vskip 4mm For symmetric configurations we have (denoting $\nu = n
- 2$)
\begin{equation}
\tx = 2^{-\nu} \te + \Delta, \quad \Delta = 4\beta^{-n} (1 -
\beta). \label{txs}
\end{equation}
Thus, the points representing configurations with given $\beta$
lie on a straight line in the $(\te, \tx)$ plane with the slope
$2^{-\nu}$ and the shift along the $\tx$-axis $\Delta$. The shift
is negative for $\beta > 1$ and reaches minimum with the value
$\Delta_0 = - (4/n_-) (n_-/n)^n$ at $\beta_0 = n/n_-$. If $n \gg
1$, the constants $\beta_0$ and $\Delta_0$ are close to 1 and 0
respectively, $\beta_0 \doteq 1 + 1/n$ and $\Delta_0 \doteq -
4/(en)$.

\vskip 4mm Consider now asymmetric configurations. For $\te$ as a
function of $y$ we have
\begin{equation}
\te = - (1/n_-) \beta^{-n} \psi^{-1}, \quad \psi = (y\bar y)^2
\phi, \label{te}
\end{equation}
and $\tx$ as a function of $y$ is given by equation (\ref{tx})
with the above expression inserted for $\te$. For given parameter
$\beta$, this defines the line $b^{(-)}$ with the slope
\begin{equation*}
\frac {d\tx}{d\te} = \frac {\tx'}{\te'} = \frac {\partial_y \tx +
\partial_{\tte} \tx \mbox{\hskip 0.8mm$\te'$}}{\te'} =
\partial_{\tte} \tx = y^{n_-} + \bar y^{n_-}.
\end{equation*}
For $y = 1/2$ the slope coincides with that for symmetric
configurations, $d\tx/d\te = 2^{-\nu}$, and for decreasing $y$ it
increases, approaching 1 as $y$ goes to 0. The parameter $\te$ at
the same time goes to $-\infty$. However, this does {\it not} mean
that as $\te$ decreases, the line $b^{(-)}$ approaches straight
line under the angle 45$^{\circ}$ to the $\te$-axis. Denote $z =
y\bar y$. For $z \ll 1$ we have $\te \doteq  - (1/n_-) \beta^{-n}
z^{-2}$ and
\begin{equation*}
\tx \doteq \te (1 - n_- z) + \beta^{-n} z^{-1} \doteq \te +
2n_-^{1/2} \beta^{-n/2} (-\te)^{1/2},
\end{equation*}
so that $\te_+ \doteq 2\te$, $\te_- \doteq -2n_-^{1/2}
\beta^{-n/2} (-\te)^{1/2}$ and the line $b_{lim}$ to which
$b^{(-)}$ converges is, just as for photon decay, an upper half of
a parabola with the axis on the lower diagonal,
\begin{equation}
\te_+ = - \frac {\beta^n}{2n_-} \te_-^2, \quad \te_- < 0.
\label{acol}
\end{equation}
The true limit lines are shifted along the axis $\eta_-$ to the
left the more the closer $\beta$ to 0. In particular, for $n = 3$
the lines $b^{(-)}$ as well as $b_{lim}$ are of the form $\te =
-(1/8) \beta^3 (\te_- - 4\beta^{-2})^2$, $\te_- < -4\beta^{-3}(2 -
\beta)$, and after a simple algebra we find that their shifts
along the axes $\eta_-$ and $\eta_+$ are $\Delta \te_- =
-2\beta^{-3}(1 - 2\beta)$ and $\Delta \te_+ = \beta^{-3}(1 -
4\beta)$.

\vskip 4mm The lines of asymmetric configurations are attached to
the lines of symmetric configurations with the same $\beta$ at the
``line of matching points'' $b_{m. p.}$, given parametrically as
\begin{equation}
\te =  - 2^{n_+} \beta^{-n}/(n_- \nu), \quad \tx = 4\beta^{-n} [1
- 2/(n_- \nu) - \beta], \label{lim}
\end{equation}
where $n_+ = n + 1$. The line starts at the origin, touches the
lowest line of symmetric configurations at the point $(\te_0,
\tx_0) = (\te, \tx)\big|_{\beta = \beta_0}$ in the lower left
quadrant, and then its behavior depends on the value of $n$: for
$n = 3$ it falls down monotonically, while for $n > 3$ it
eventually stops and starts to rise. If $n \gg 1$, the point
$(\te_0, \tx_0)$ is located far from the origin just under the
$\te$ axis, $\te_0 \doteq -2^{n_+}/(en^2)$ and $\tx_0 \doteq
-4/(en)$.

\vskip 4mm One would expect that the line of asymmetric
configurations will proceed from the starting point at $\te =
\te_c$ (the value of $\te$ at the line $b_{m. p.}$) towards
smaller $\te$, falling down with increasing slope. However, such
behavior is observed only if the function $\psi$ rises
monotonically with $y$ for $y < 1/2$, or equivalently, with $z$
for $z < 1/4$. As it turns out, this is the case only if $n \le
7$. For $n = 3, 4$ the function $\psi$ equals $z^2$, hence it
rises monotonically for all $z > 0$, but for greater $n$ it
acquires a maximum that shifts with increasing $n$ towards smaller
$z$, until it falls below 1/4. This happens at $n = 8$, when $\psi
= z^2(1 - 4z + 3z^2)$ and $\psi = \psi_{max}$ at $z = (3 -
\sqrt{3})/6 = 0.211$. The maximum then shifts further, down to $y
\doteq 2n^{-1}$ for $n \gg 1$. Such behavior means that the line
$b^{(-)}$ has a cusp at some $\te_m > \te_c$; as $y$ decreases, it
first rises towards greater $\te$, and only after $\te$ reaches
the value $\te_m$ it turns back and starts to fall down.

\vskip 4mm The extremum of the parameter $\beta$ as a function of
$y$ is minimum if $\beta'' = - \partial^2_y \tx/\partial_\beta \tx
> 0$. With the expression (\ref{tx}) for $\tx$, we obtain for symmetric
configurations
\begin{equation*}
\beta'' = 2^{-\nu} \nu \beta^{n_+} \frac {\te -\te_c}{\beta_0 -
\beta},
\end{equation*}
and for asymmetric configurations
\begin{equation*}
\beta'' = \frac 14 \beta^{n_+} \frac {z^{-2} d\psi/dz} {(4z)^{-1}
\beta_0 - \beta} (-\te) (y - \bar y)^2.
\end{equation*}
We want to construct lines in the $(\te, \tx)$ plane at which the
lower threshold for $e^+ e^-$ creation equals $\beta k_{LI}$. The
lines, which we will denote by $b$, must satisfy $\beta' = 0$ and
$\beta'' > 0$, and if two such lines with different values of
$\beta$ cross at the given point in the $(\te, \tx)$ plane, we
must chose the one with the less $\beta$.

\vskip 4mm Suppose first that $\beta < \beta_0$. Determining the
line $b$ is straightforward if $n \le 7$. For such $n$ it holds
$d\psi/dz > 0$ for all $z < 1/4$, therefore $\beta'' > 0$ along
the whole line of asymmetric configurations. Note also that
$\partial_\beta \tx = - 4n_- \beta^{n_+}[(4z)^{-1} \beta_0 -
\beta] < 0$ for all $z < 1/4$, hence the lines of asymmetric
configurations do not intersect. Furthermore, the line of
symmetric configurations has $\beta'' > 0$ for all $\te > \te_c$,
that is, all the way up from the matching point with the line of
asymmetric configurations to infinity. Thus, if we denote the part
of the line of symmetric configurations with $\te > \te_c$ by
$b^{(+)}$, the line $b$ is the union of $b^{(-)}$ and $b^{(+)}$.
The analysis is a bit more tricky if $n > 7$. The line $b^{(-)}$
is then composed of two parts, the part $b^{(-)}_I$ which goes
from the point $\te = \te_c$, where it matches the line of
symmetric configurations, to the cusp at $\te = \te_m$, and the
part $b^{(-)}_{II}$ which goes from the cusp to infinity. Along
the former part it holds $\beta'' < 0$ and along the latter part
it holds $\beta'' > 0$. Furthermore, since the derivative
$d\tx/d\te$ increases as we move from the matching point through
the cusp to infinity, the lines $b^{(-)}_I$ and $b^{(-)}_{II}$ are
convex and concave respectively; and since the line $b^{(-)}_I$ is
tangential to the line of symmetric configurations at the matching
point, the cusp is located above that line. Thus, the lines
$b^{(-)}_{II}$ and $b^{(+)}$, at which both conditions $\beta' =
0$ and $\beta'' > 0$ are satisfied, intersect at some point $\te =
\te_i$ between the matching point and the cusp. Denote the line
composed of $b^{(-)}_{II}$ and $b^{(+)}$ by $b_0$ and consider two
neighboring lines $b_0$ and $\tilde b_0$ with $\tilde \beta <
\beta$. The line $\tilde b^{(+)}$ (upper part of $\tilde b_0$)
crosses the line $b^{(-)}_{II}$ (left part of $b_0$) at $\te >
\te_i$, that is, above the point of intersection of the lines
$b^{(-)}_{II}$ and $b^{(+)}$; and the line $\tilde b^{(-)}_{II}$
(left part of $\tilde b_0$) crosses the line $b^{(+)}$ (upper part
of $b_0$) at $\te < \te_i$, that is, left to the point of
intersection of the lines $b^{(-)}_{II}$ and $b^{(+)}$. At the
crossing points, the lower threshold of $e^- e^+$ annihilation is
located at the curve with lower $\beta$, which is $\tilde b_0$. We
can see that in order to obtain the line $b$ we must remove from
the line $b_0$ the part of $b^{(-)}_{II}$ above the point of
intersection, as well as the part of $b^{(+)}$ left to the point
of intersection. Thus, $b$ is the union of the parts of the lines
$b^{(-)}_{II}$ and $b^{(+)}$ going from infinity to the point of
intersection and from the point of intersection back to infinity.

\vskip 4mm Suppose now that $\beta > \beta_0$. The part of the
line of symmetric configurations complementary to $b^{(+)}$ does
not contribute to $b$ because it has greater value of $\beta$ than
the line of asymmetric configurations crossing it at any given
point. Thus, we are left with the line of asymmetric
configurations $b^{(-)}$, or rather its part $b^{(-)}_{down}$
which is cut either at the line $b^{(+)} (\beta_0)$, that is, at
$z = z_1$ such that $\tx_1 = 2^{-\nu} \te_1 + \Delta_0$, or at the
point where the line $b^{(-)}$ intersects the neighboring line
$\tilde b^{(-)}$, that is, at $z = z_2 = (1/4) \beta_0/\beta$,
whichever point comes first as we follow the line from large
negative $\te$ to $\te = \te_c$. (The cut at $z = z_2$ is
necessary since at smaller $z$ it holds $\beta'' < 0$.) For $n =
3$, the cut occurs at the former point (it holds $z_1^{-1} =
4\big[1 + u\sqrt{(3 + u)/2}\big]$, $u = \beta/\beta_0 - 1$, so
that $z_1^{-1} > z_2^{-1} = 4(1 + u)$), and we will assume that
the same is true for $n > 3$, because even if it was not, the form
of the allowed region discussed further would stay qualitatively
the same.

\vskip 4mm Suppose, following \cite{jac}, that the lower threshold
of $e^- e^+$ annihilation lies between $k_{LI}$ and $2k_{LI}$. The
allowed region in the $(\te, \tx)$ plane is then an infinite band
between the line $b(1)$ and the union of the lines $b(2) =
b^{(-)}_{down}(2)$ and $b^{(+)} (\beta_0)$ cut at the end point of
$b(2)$. The band, if we follow it from large positive to large
negative values of $\te$, is first straight, keeping its width and
tilted downwards with the slope $2^{-\nu}$, and then it widens and
bends downwards, becoming parabolic with the axis on the lower
diagonal as $\te \to -\infty$. While still straight, the band
touches the origin from below.

\vskip 4mm The allowed region in the complete theory, by which we
mean the theory of the three processes considered here, is an
intersection of the band we have just constructed with the
infinite wedge we have constructed earlier. To see how this region
looks like, we must pass from the dimensionless parameters $(\hat
\eta, \hat \xi)$, $(\tilde \eta, \tilde \xi)$ and $(\te, \tx)$ to
the dimensional parameters $(\eta, \xi)$; that is, we must
multiply the parameters $(\hat \eta, \hat \xi)$ by $A =
m^2/(p_{max}^{obs})^n$, the parameters $(\tilde \eta, \tilde \xi)$
by $B = m^2/(k_{max}^{obs})^n$ and the parameters $(\te, \tx)$ by
${\cal B} = \omega_0^n/m^{2n_-} = m^2/k_{LI}^n$. For the momenta
appearing in these expressions, let us adopt the values used in
\cite{jac}, namely $p_{max}^{obs} = k_{max}^{obs} =$ 50 TeV and
$k_{LI} =$ 10 TeV (corresponding to $\omega_0 =$ 25 meV). In
Planck units, the constants $A$, $B$ are
\begin{equation}
\mbox{$A$, $B$} \doteq \Big(\frac {0.5 \mbox{ MeV}}{50 \mbox{
TeV}}\Big)^2 \Big(\frac {10^{19} \mbox{ GeV}}{50 \mbox{
TeV}}\Big)^\nu = 2^\nu \times 10^{- 16 + 14\nu} = 0.02, 4 \times
10^{12}, \ldots \mbox{ for } n = 3, 4, \ldots, \label{15}
\end{equation}
and the constant $\cal B$ is greater than the constants $A$, $B$
by the factor $5^n$. Using equations (\ref{adec}) and
(\ref{acol}), we find that the width of the inner boundary of the
allowed region for photon collision, when considered far from the
origin, is greater than the width of the outer boundary of the
allowed region for photon decay by the factor $(5/2)^n = 15.6, 39,
\ldots$ for $n = 3, 4, \ldots$ (The width of a parabola is defined
as the distance between opposite points at the level of focus, $d
= k^{-1}$ for $y = kx^2$.) Thus, the bent segment of the former
region lies far to the left of the latter region, deep in the
forbidden part of the $(\eta, \xi)$ plane.

\vskip 4mm The allowed region for the three processes considered
here is depicted in fig. 6.
\begin{figure}[ht]
\centerline{\includegraphics[height=5.4cm]{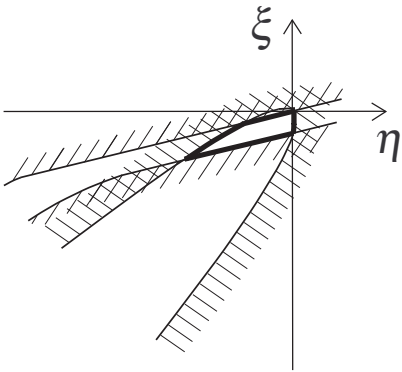}}
 \caption{\small Allowed region in the complete theory}
\end{figure}
The region, delimited by heavy line, is an intersection of two
regions delimited by light lines, the allowed region for the two
one-particle processes (the wedge) and the allowed region for the
two-particle process (the bowed band). As we can see, the region
has the form of a tilted trapezoid-like strip, with the upper
right vertex close to the origin and the right-hand side close to
the $\xi$ axis. This is just the kind of behavior that has been
observed earlier in the case $n = 3$, see fig. 8 in \cite{jac}.

\section{Conclusion}

In a theory with dispersion relations (\ref{1}) one would expect
$n$ to be small, say, 2, 3 or 4, and $\xi$ and $\eta$ to be of
order $m_{Pl}^{-(n - 2)}$, where $m_{Pl}$ is Planck mass. To see
how far the theory can be stretched, we have supposed that $n$ as
well as $\xi$ and $\eta$ can be arbitrary, requiring just that the
dispersion relations do not contradict observational data. From
the fact that the highest energy of electrons and photons
encountered in observations is by many orders of magnitude less
than the Planck mass it follows that large values of $n$ bring in
large values of $\xi$ and $\eta$: as seen from equation
(\ref{15}), $\xi$ and $\eta$ are typically of order $10^{-16}
\times (2 \times 10^{14})^{n - 2} m_{Pl}^{-(n - 2)}$, so that they
rise steeply with $n$ when expressed in Planck units. The
corresponding mass scale is 50 $m_{Pl}$ for $n = 3$, it falls down
to $5 \times 10^{-7} m_{Pl}$ for $n = 4$, and as we increase $n$,
it continues to decrease, approaching gradually the value $5
\times 10^{-15} m_{Pl}$ (maximum energy available in
observations). Of course, the parameters $\xi$ and $\eta$ do not
need to be from the bulk of the allowed region, we can assume that
they are from a tiny patch around the origin. That would push the
mass scale towards $m_{Pl}$, however, we should then come to terms
with the fact that the deviation from standard dispersion
relations will not be observed any soon.

\vskip 4mm Two objections can be raised against large values of
$n$: there is no reason that in the Taylor expansion of energy as
a function of momentum a lot of terms is skipped before the
expansion starts; and it does not seem plausible for the future
theory of quantum gravity, whatever it will look like, to lead to
mass scales that are substantially less than $m_{Pl}$. We did not
attempt to propose a theory in which $n$ would be large and $\xi$
and $\eta$ would be much greater than $m_{Pl}^{-(n - 2)}$.
Instead, our aim was to determine, in the spirit of
quantum--gravity phenomenology, how the observational constraints
would look in a theory with large $n$, knowing in advance that we
will need also large $\xi$ and $\eta$ in order to be able to
actually observe the effect of the additional term in dispersion
relations. We have found out, by analyzing the three main
processes determining the boundaries of the allowed region in the
$(\eta, \xi)$ plane, that the region is similar in shape to that
obtained in \cite{jac} for $n = 3$, and is stretched by a factor
$2 \times 10^{14} m_{Pl}^{-1}$ each time we increase $n$ by unity.

\enddocument